\documentclass[preprint, secnumarabic, amssymb, amsmath, aip, rsi, groupedaddress, frontmatterverbose]{revtex4-1}

\usepackage{graphicx,color}
\DeclareGraphicsExtensions{.jpeg, .png, .pdf, .eps}

\usepackage{docs}
\usepackage{bm}
\expandafter\ifx\csname package@font\endcsname\relax\else
 \expandafter\expandafter
 \expandafter\usepackage
 \expandafter\expandafter
 \expandafter{\csname package@font\endcsname}
\fi
\begin{document}

\title{Optimization of a charge-state analyzer for ECRIS beams}
\author{S.~Saminathan,\footnote{Present address: TRIUMF, 4004 Wesbrook Mall, Vancouver, BC, V6T 2A3, Canada.} J.P.M.~Beijers,\footnote{Author to whom correspondence should be addressed. Electronic mail: beijers@kvi.nl.} H.R.~Kremers, V.~Mironov, J.~Mulder, and S.~Brandenburg}
\affiliation{Kernfysisch Versneller Instituut, University of Groningen, Zernikelaan 25, 9747 AA Groningen, The Netherlands}
\date{\today}
\pacs{29.25.Ni, 41.75.Ak, 41.85.Ar, 41.85.Gy, 41.85.Qg}
\keywords{ECR ion source, Ion beam transport, Aberration compensation}
\maketitle

\section{Abstract}
A detailed experimental and simulation study of the extraction of a 24~keV He$^+$ beam from an ECR ion source and the subsequent beam transport through an analyzing magnet is presented. We find that such a slow ion beam is very sensitive to space-charge forces, but also that the neutralization of the beam's space charge by secondary electrons is virtually complete for beam currents up to at least 0.5~mA. The beam emittance directly behind the extraction system is 65~$\pi$~mm~mrad and is determined by the fact that the ion beam is extracted in the strong magnetic fringe field of the ion source. The relatively large emittance of the beam and its non-paraxiality lead, in combination with a relatively small magnet gap, to significant beam losses and a five-fold increase of the effective beam emittance during its transport through the analyzing magnet. The calculated beam profile and phase-space distributions in the image plane of the analyzing magnet agree well with measurements. The kinematic and magnet aberrations have been studied using the calculated second-order transfer map of the analyzing magnet, with which we can reproduce the phase-space distributions of the ion beam behind the analyzing magnet. Using the transfer map and trajectory calculations we have worked out an aberration compensation scheme based on the addition of compensating hexapole components to the main dipole field by modifying the shape of the poles. The simulations predict that by compensating the kinematic and geometric aberrations in this way and enlarging the pole gap the overall beam transport efficiency can be increased from 16 to 45\%.

\section{Introduction}
The operational experience with ECR ion sources at heavy ion accelerators shows that in general significant losses occur in the low energy beam transport system between the ion source and the accelerator. As an example, the 25~m long beam line between the KVI Advanced Electron Cyclotron
Resonance (KVI-AECR) ion source~\cite{kremers1} and the AGOR-cyclotron~\cite{brandenburg1} has a typical transmission of  about $16\%$~\cite{thesis}. Similar values have been reported by other laboratories~\cite{stetson}. Various hypotheses for this low transmission have been proposed, mostly focusing on the large and convoluted phase-space distributions of the beams extracted from ECR sources and the angular momentum of the beams caused by the combination of electric and magnetic fields in the extraction region of the source~\cite{stetson}. However, not much convincing evidence for these hypotheses has been presented up to now.

In order to better understand and possibly improve the beam transport efficiency we have performed a simulation and experimental study of the beam extraction from the KVI-AECR ion source and transport through the low-energy beam line~\cite{thesis}. This study shows that the relatively large emittances of ion beams extracted from ECR ion sources, in combination with the relatively small vertical gap of the $110^{\circ}$ analyzing magnet, cause large second-order kinematic and magnet aberrations resulting in beam losses and a five-fold increase of the effective beam emittance. The emittance blowup in the analyzing magnet leads to large beam losses further down the beam line and is the main cause of the poor transmission.

In addition, we describe a method to mitigate the emittance blowup in the analyzing magnet and improve the beam transport efficiency. This can be achieved by increasing the magnet gap and suitably modifying the shape of the pole faces. We have done all simulations and measurements discussed in this paper for the case of a mono-component 24~keV He$^+$ beam extracted from the KVI-AECR ion source. In this way measurements and simulations can be unambiguously compared both before and behind the analyzing magnet. Quoted values of beam emittances always refer to the 4-rms emittance as defined by Lapostolle, which encloses 90\% of the beam\cite{lapostolle}.

This paper is organized as follows. In section~\ref{sec:trans} we briefly describe the KVI-AECR ion source, the first section of the low-energy beam transport line and a transport simulation of a He$^{+}$ beam through the analyzing magnet. A detailed study of the ion-optical aberrations of
the analyzing magnet based on trajectory simulations and a second-order analysis of the beam transport through the magnet is presented in section~\ref{sec:abber}. Then in section~\ref{sec:abber_corr} a method is described to compensate the ion-optical aberrations of the analyzing magnet by adding hexapole components to its main dipole field. The paper closes with conclusions and outlook in section~\ref{sec:conlook}.

\section{\label{sec:trans}Extraction and transport of a He$^{+}$ beam}
\begin{figure}[tb!]
\centering
\includegraphics[width=7.5cm]{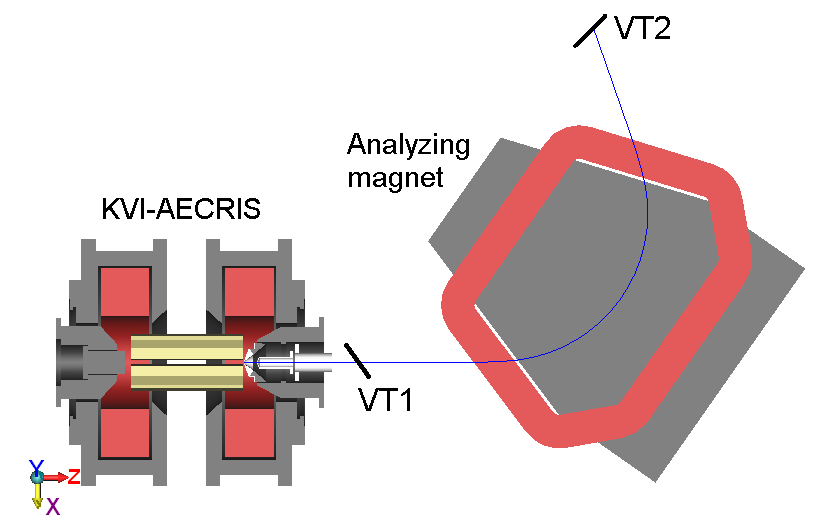}
\caption{\label{fig:aecrm110}AECR ion source and first section of the low-energy beam transport line. The locations of the two viewing screens are
indicated as VT1 and VT2.}
\end{figure}

The KVI-AECR ion source and the first section of the low-energy beam transport line including the $110^{\circ}$ analyzing magnet are schematically shown in Figure~\ref{fig:aecrm110}. The ion source is a 14~GHz electron cyclotron resonance ion source using two room temperature solenoids and an open NdFeB permanent magnet hexapole to create a min-$B$ plasma trap. A more detailed description of the AECR ion source can be found in Ref.~\cite{kremers2}. The electrostatic extraction system is an accel-decel lens consisting of a plasma electrode with a 8~mm diameter extraction aperture followed by shielding and ground electrodes. The shielding electrode is typically biased at $-300$~V with respect to the ground electrode to prevent secondary electrons, produced downstream of the shielding electrode, from being back accelerated into the ion source and thus to maintain optimum space-charge compensation in the downstream beam line.

Following the extraction system is an unclamped double-focusing dipole magnet (hereafter denoted as analyzing magnet) which bends the ion beam over $110^{\circ}$ to select ions with a given charge-to-mass ratio. The analyzing magnet has a geometrical acceptance of 120~mm in the horizontal and 60~mm in the vertical direction, determined by the size of the vacuum chamber. The bending radius of the magnet is 400~mm. Its pole faces have a tilt angle of $37^{\circ}$ to obtain simultaneous imaging in both transverse planes. The distance from the extraction aperture of the plasma electrode to the effective field boundary (EFB) of the magnet is 682~mm and the distance between the EFB and the image plane is 374~mm. Following the analyzing magnet the ions are transported through a series of bending and focusing elements to the injection system of the AGOR cyclotron.
Two BaF$_2$ viewing screens can be inserted into the beam line to measure the spatial distribution of the ion beam, one located at a distance of 334~mm to the effective field boundary (EFB) of the analyzing magnet denoted as VT1 and another close to the image plane of the analyzing magnet
denoted as VT2. In addition we have measured the transversal phase-space distribution behind the analyzing magnet at the location of VT2 with a pepperpot emittance meter~\cite{kremers3}. These measurements have been used to benchmark the simulations.

The simulation of the extraction and transport of the He$^+$ beam consists of two parts. First the phase-space distribution of the He$^+$ ions in the plane of the plasma electrode is calculated using a dedicated Particle In Cell-Monte Carlo Collision (PIC-MCC) code~\cite{mironov}. This code
simulates the ion dynamics in the ECR plasma taking into account the full three-dimensional field geometry of the min-$B$ trap. The phase-space coordinates of the extracted ions at the position of the plasma electrode are then used as initial conditions to calculate the ion trajectories through the extraction system, drift space and analyzing magnet. For these trajectory calculations we use a particle tracking code, i.e.\ the GPT code~\cite{gpt}, taking into account the three-dimensional electric and magnetic fields in the extraction system and analyzing magnet. The LORENTZ-3D code has been used to calculate these three-dimensional field distributions\cite{lorentz}. From the calculated ion trajectories we extract four-dimensional transversal phase-space distributions at the VT1 and VT2 locations and compare two-dimensional projections of these distributions with measured ones.

The effect of space charge has been studied by performing a series of beam transport simulations for a primary beam intensity of 450~$\mu$A with the degree of space-charge compensation varying from 95\% to 0\%. The simulations show a strong effect of the degree of space-charge compensation on both the spatial and phase-space distributions. The spatial distribution calculated at VT1 has 'hot' spots for a space-charge compensation of 90\% and develops a hollow core when the compensation is less than 90\%. The calculated emittance increases linearly from 65~$\pi$~mm~mrad for a fully compensated beam to 225~$\pi$~mm~mrad for a fully uncompensated beam. The measured beam images at VT1 and behind the analyzing magnet are consistent with the calculated ones having 95-100\% space-charge compensation. Therefore, all transport simulations reported in the rest of this paper have been performed assuming fully compensated ion beams.

\begin{figure}[tb!]
\centering
\includegraphics[width=12cm]{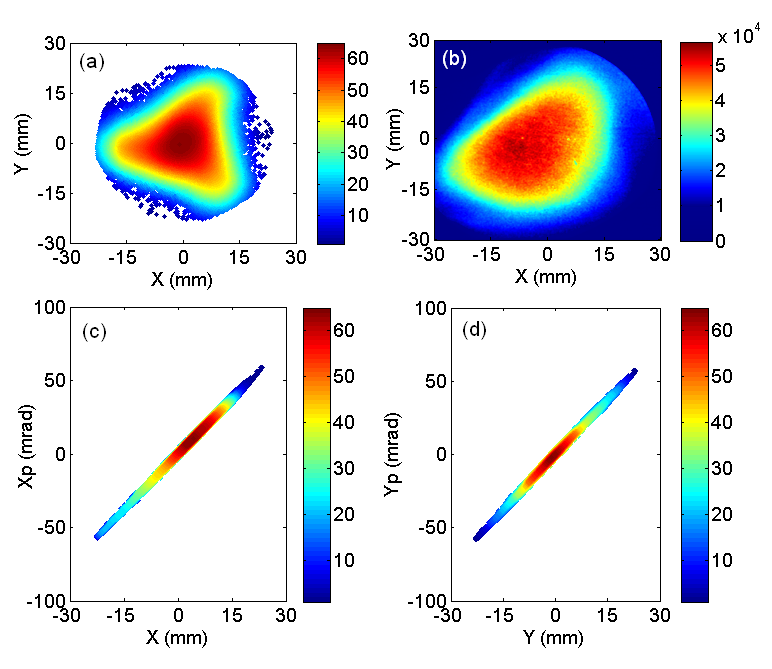}
\caption{\label{fig:he_prf_vt1}Spatial and transverse phase-space distributions of a 24~keV He$^+$ beam at location VT1. Simulated (a) and measured (b) spatial distribution. Simulated horizontal (c) and vertical (d) phase-space distribution.}
\end{figure}

Calculated spatial and phase-space distributions as well as a measured spatial image of a 24~keV He$^+$ beam at the location of VT1 are shown in Figure~\ref{fig:he_prf_vt1}. The calculated and measured spatial distributions show the characteristic triangular shape of beams extracted from ECR ion sources and agree reasonably well. The simulated transverse phase-space distributions have elliptical shapes and show that the ion beam is strongly diverging at this location. The beam emittance is $65~\pi$~mm~mrad in both transverse directions and is mainly due to the fact that the beam is extracted in the magnetic fringe field of one of the source solenoids. We checked this by repeating the extraction simulation with the magnetic fringe fields in the extraction region turned off, which indeed yielded a negligible emittance ($\approx 2~\pi$~mm~mrad).

\begin{figure}[tb!]
\centering
\includegraphics[width=12cm]{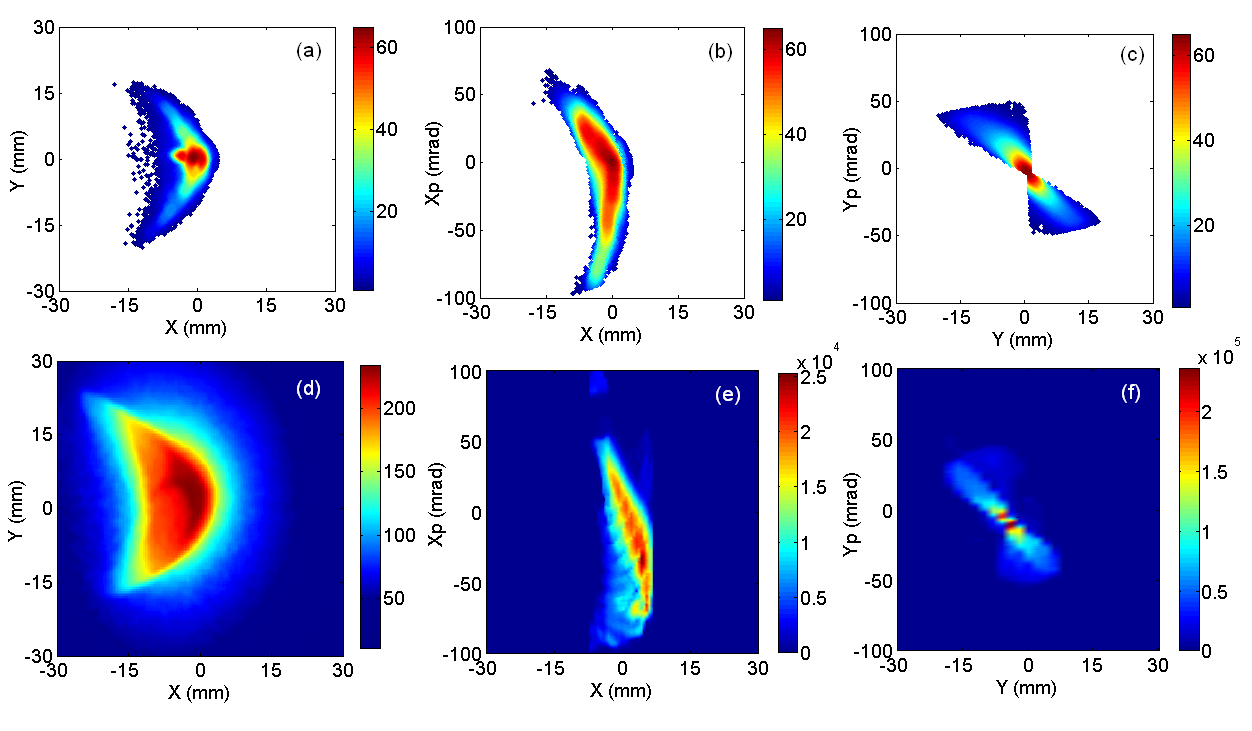}
\caption{\label{fig:he_prf_vt2}Spatial and transverse phase-space distributions of a 24~keV He$^+$ beam at location VT2. Simulated (a) and measured (d) spatial distribution, simulated (b) and measured (e) horizontal phase-space distribution, simulated (c) and measured (f) vertical phase-space distribution.}
\end{figure}

Simulated as well as measured spatial and transverse phase-space distributions of the He$^+$ beam behind the analyzing magnet at the location of VT2 are shown in Figure~\ref{fig:he_prf_vt2}. The triangular spatial distribution of the beam at the location of VT1 is distorted into a crescent-shaped distribution (Figure~\ref{fig:he_prf_vt2}a and d). This is an indication of strong aberrations, in particular second-order, of the analyzing magnet. Magnet aberrations also cause a large distortion of the horizontal and vertical phase-space distributions
at the location of VT2 as shown in Figure~\ref{fig:he_prf_vt2}b, c, e and f. They lead to a large increase of the beam emittance in both transverse planes. The simulated and measured emittances of the transported beam in the horizontal plane are $360~\pi$~mm~mrad and
$390~\pi$~mm~mrad, respectively, and in the vertical plane $240~\pi$~mm~mrad and $320~\pi$~mm~mrad, respectively. According to the simulation the beam loss during the transport from the location of VT1 to VT2 is around $25\%$, and is mainly caused by the rather small vertical aperture of the magnet. This also explains the smaller emittance in the vertical plane.

The calculated and measured distributions at the locations VT1 and VT2 agree reasonably well. This indicates that the physical mechanisms determining ion beam formation, extraction and transport are well represented in our simulation models and that the ion beam is (almost) completely space-charge compensated. The beam emittance behind the analyzing magnet is significantly larger than the acceptance of the AGOR injection system, which is 140~$\pi$~mm~mrad, thus leading to large beam losses. In order to improve the transport efficiency of the low-energy beam transport line we have to minimize emittance growth in the analyzing magnet.

\section{\label{sec:abber}Transport properties of the magnet}
\begin{figure}[tb!]
\centering
\includegraphics[width=12cm]{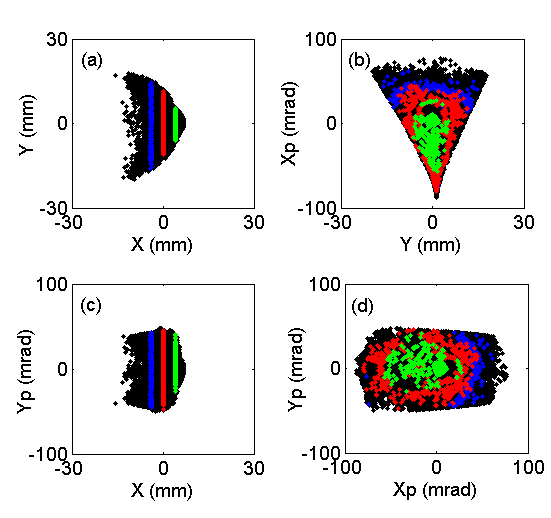}
\caption{\label{fig:He1_m110_vt2_px_corr}Simulated projections of phase-space distributions at location VT2 behind the analyzing magnet. The colored bands indicate He$^+$ ions centered at horizontal positions $x=0$ and $\pm 5$~mm.}
\end{figure}
\begin{figure}[tb]
\centering
\includegraphics[width=10cm]{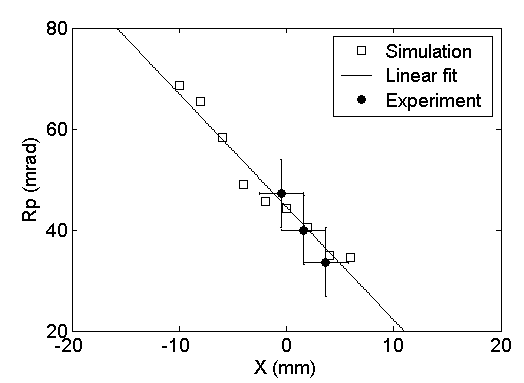}
\caption{\label{fig:He_vt2_FRp}Simulated ($\Box$) and measured (\textbullet) transverse angles $R_p$ as a function of the horizontal position $x$ in the image plane of the analyzing magnet at location VT2.}
\end{figure}

The measurements and simulations presented in the previous section indicate large second-order effects in the phase-space distributions behind the analyzing magnet and correlations between the horizontal and vertical phase-space coordinates of the beam particles. This is illustrated in Figure~\ref{fig:He1_m110_vt2_px_corr} where various simulated projections of the 4d transverse phase-space are plotted at the location of VT2 behind the analyzing magnet. The colored bands indicate groups of particles which are centered at the horizontal positions $x=0$ and $\pm 5$~mm. Figure~\ref{fig:He1_m110_vt2_px_corr}d suggests that a useful quantitative measure of correlations between phase-space coordinates is provided
by the transverse angle $R_p$ defined as
\begin{equation}
R_p = \sqrt{x_p^2+y_p^2}\;, \label{equ:Rp}
\end{equation}
The angle $R_p$ is proportional to the transverse momentum of the ions. Figure~\ref{fig:He1_m110_vt2_px_corr}d indicates a strong correlation between $R_p$, i.e.\ distance to the origin, and the $x$-coordinates of the ions in the focal plane of the analyzing magnet (color). In Figure~\ref{fig:He_vt2_FRp} both simulated and measured $R_p$ values are plotted as a function of the $x$-coordinates of the ions at location VT2 behind the analyzing magnet, showing a linear correlation between $x$ and $R_p$. The experimental data points have been
measured with a pepperpot emittance meter. A fixed horizontal offset has been added to the experimental data points since the horizontal position of the pepperpot has not been calibrated absolutely.

The large second-order correlations observed in both simulations and measurements are essentially caused by non-paraxiality of the beam (kinematic aberrations) and second-order geometric aberrations in the analyzing magnet. Initial correlations in the beam due to the ECR source can be neglected. We checked this by performing transport simulations for a hypothetical rotationally symmetric and uniform He$^+$ beam with an emittance of $60~\pi$~mm~mrad at location VT1 behind the extraction system of the ECR source. Tracking this beam through the analyzing magnet shows that the simulated phase-space distributions and emittance values at location VT2 behind the analyzing magnet are essentially identical to those of an actual beam extracted from the ECR source.

\begin{figure}[tb!]
\centering
\includegraphics[width=12cm]{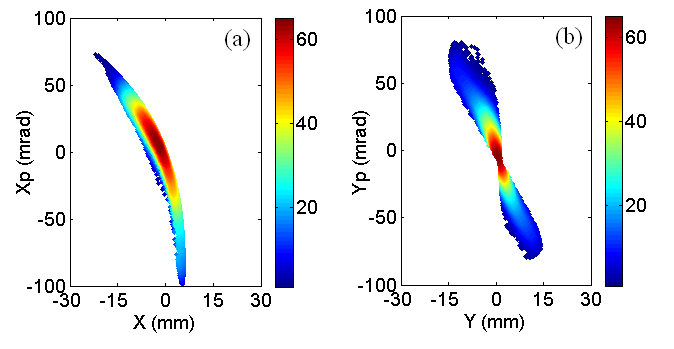}
\caption{\label{fig:He1_cosy_fit_vt2}Second-order mapping of the phase-space distribution of a 24~keV He$^+$ beam at location VT1 and shown in Figure~\ref{fig:he_prf_vt1}c and d to location VT2 behind the analyzing magnet using equation~(\ref{eq:x2nd}). The expansion coefficients have been
calculated with COSY INFINITY.}
\end{figure}

In order to investigate the aberrations of the analyzing magnet we have calculated its second-order transfer matrix describing the mapping of phase-space coordinates $\theta$, with $\theta=x,y,x'$ or $y'$, from VT1 to VT2 and which can be written as\cite{brown}
\begin{equation}
\begin{split}
   \theta_1 = &(\theta|x)x_0+(\theta|x')x'_0+(\theta|y)y_0+(\theta|y')y'_0+(\theta|xx)x_0^2+\\
   &+(\theta|xx')x_0x'_0+(\theta|x'x')x_{0}'^2+(\theta|xy)x_0y_0+(\theta|x'y)x'_0y_0+(\theta|xy')x_0y'_0+\\
   &+(\theta|x'y')x'_0y'_0+(\theta|yy)y_0^2+(\theta|yy')y_0y'_0+(\theta|y'y')y_{0}'^2\;,
\end{split}
\label{eq:x2nd}
\end{equation}
with subindexes 0 and 1 indicating locations VT1 and VT2, respectively. The first- and second-order coefficients in equation~(\ref{eq:x2nd}) have been calculated with the COSY INFINITY code, using an internal model (model 2) for the fringe field of the analyzing magnet\cite{cosy}. Equation~(\ref{eq:x2nd}) has been used to map the 4d phase-space distribution at location VT1, shown in Figure~\ref{fig:he_prf_vt1}c and d, to location VT2. The result is shown in Figure~\ref{fig:He1_cosy_fit_vt2} and compares very well with the phase-space distributions obtained by full particle tracking shown in Figures~\ref{fig:he_prf_vt2}e and f. We have also extended the calculated transfer map to third order, but this did not change the results. The transport properties of the analyzing magnet are thus indeed dominated by its second-order aberration coefficients.

The parabolic shape of the image at location VT2 is mainly due to the second-order coefficients $(x|yy)$ and $(x|y'y')$. Particles with large values of $y$ and/or $y'$ at location VT1 are mapped to large values of $y$ at location VT2 behind the analyzing magnet. This results in a crescent shape image in the $xy$-plane. The $(x|x'x')$ term causes the banana-shaped phase-space distribution in the $xx'$-plane. The bow-tie shape phase-space distribution in the $yy'$-plane is mainly due to the second-order terms in the $y$-column of the transfer map, equation~(\ref{eq:x2nd}). All nonlinear terms together cause a five-fold increase in the emittance of the He$^+$ beam transported through the analyzing magnet. This leads to significant beam losses in the downstream beam transport and injection into the AGOR cyclotron.

\section{\label{sec:abber_corr}Second-order correction}
The analysis described in the previous section has shown that the relatively small vertical gap and large second-order aberration coefficients of the analyzing magnet cause significant beam losses and emittance blowup of the transported beam. The transport properties of the magnet can be improved by remedying these shortcomings, i.e.\ enlarging the magnet gap and minimizing and/or compensating the second-order aberrations. Simulations show that the beam losses on the pole face of the magnet can simply be prevented by increasing the magnet gap from $67$ to $110$~mm. Magnet
aberrations can be minimized by keeping the beam inside the magnet gap vertically narrow, e.g.\ by using an extra vertical focusing element between the ECR extraction
system and the analyzing magnet. However, in our case this is difficult because of lack of space. We have instead investigated the possibility to compensate both the kinematic and geometric second-order magnet aberrations by adding suitable hexapole components to the main dipole field of the analyzing magnet~\cite{camplan,wollnik}. This method was first used to improve the transport of intense multiply-charged ion beams extracted from ECR ion sources by Leitner \emph{et al.}~\cite{leitner}.
\begin{figure}[tb!]
\centering
\includegraphics[width=10cm]{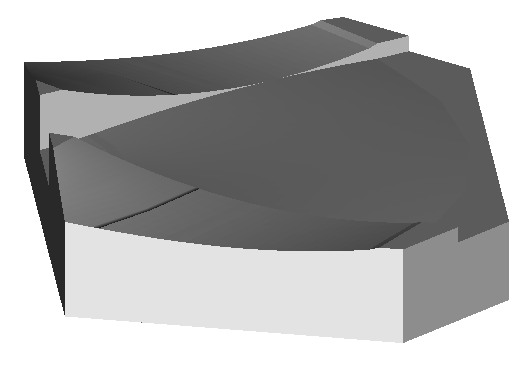}
\caption{\label{fig:modified_pole}Modified pole surface of the analyzing magnet to compensate its second-order aberrations.}
\end{figure}

\begin{figure}[tb!]
\centering
\includegraphics[width=10cm]{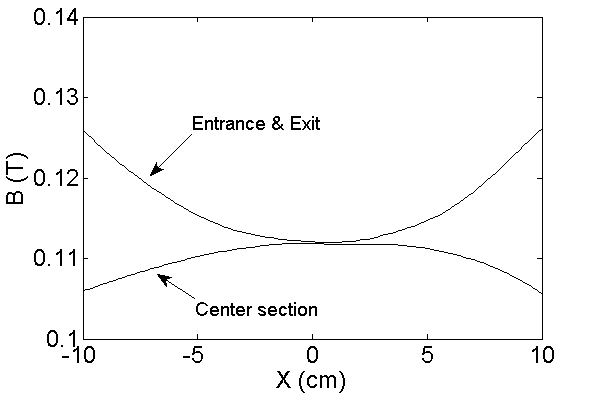}
\caption{\label{fig:Bfield_ap5p2}Optimum magnetic field profile in the middle of the entrance (exit) and central sections as a function of the horizontal distance from the central trajectory.}
\end{figure}

We have investigated the effects of modifying the pole surfaces of the entrance and exit sections of the analyzing magnet such that a quadratically increasing field is obtained to correct the hexapole component in the vertical plane. The central part of the pole surface is shaped in such a way to obtain a quadratically decreasing magnetic field in order to correct the hexapole component in the horizontal plane and to keep the total magnetic-field integral along the ion trajectories the same as before. The modified pole surface of the analyzing magnet is shown in Figure~\ref{fig:modified_pole}. We used COSY INFINITY to quickly estimate the required hexapole strengths and then the LORENTZ3D code to fine tune the shape of the pole face on the basis of trajectory calculations. As before, the initial conditions for the transport simulations of the modified magnet are taken from the calculated phase-space distribution at location VT1 (see Figure~\ref{fig:he_prf_vt1}c and d) and we assumed a fully-compensated
He$^+$ beam, i.e.\ space-charge forces have been neglected. After a few iterations we arrived at the optimum magnetic field profile shown in Figure~\ref{fig:Bfield_ap5p2}.

\begin{figure}[tb!]
\centering
\includegraphics[width=10cm]{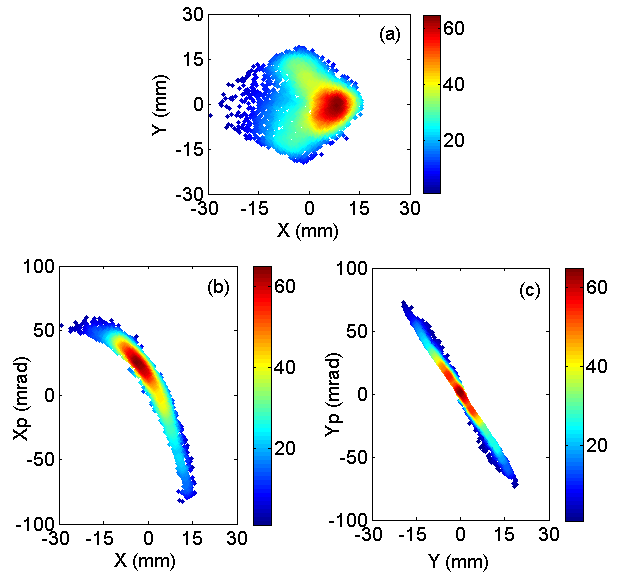}
\caption{\label{fig:corrected_prf}Simulated spatial distribution for a fully space-charge compensated He$^{+}$ beam at the location of VT2 behind the modified analyzing magnet (a). Simulated horizontal (b) and vertical (c) emittance plots at the same location.}
\end{figure}

\begin{figure}[tb!]
\centering
\includegraphics[width=12cm]{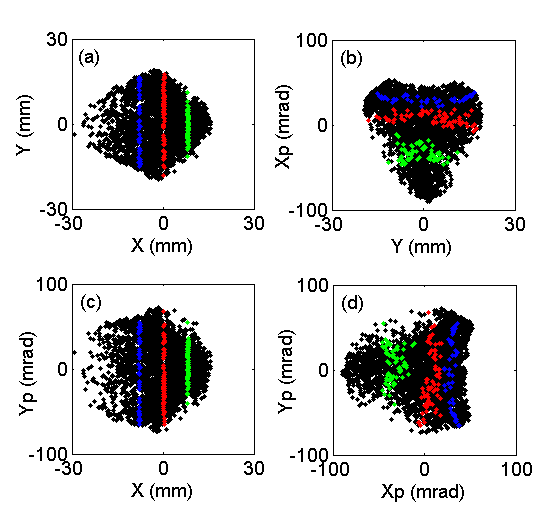}
\caption{\label{fig:He1_ap5p2_vt2_corr}Simulated projections of phase-space distributions at location VT2 behind the modified analyzing magnet. The colored bands indicate He$^+$ ions centered at horizontal positions $x=0$ and $\pm 5$~mm. Compare with Figure~\ref{fig:He1_m110_vt2_px_corr}.}
\end{figure}

The calculated beam profile and emittance distributions at the location of VT2 for the modified pole surface are shown in Figure~\ref{fig:corrected_prf}. According to the simulations the full beam is transported to the location of VT2 without losses and the horizontal and vertical
emittances are reduced by a factor of two compared to those for the uncorrected magnet. We have also recalculated Figure~\ref{fig:He1_m110_vt2_px_corr} for the modified analyzing magnet, the result is plotted in Figure~\ref{fig:He1_ap5p2_vt2_corr}. Comparison with Figure~\ref{fig:He1_m110_vt2_px_corr} shows that the correlations that existed in the various phase-space projections are almost completely removed by optimizing the magnet. As can be seen in Figure~\ref{fig:corrected_prf}(a) and (b) a small ($\approx 10\%$) fraction of the beam
particles is deflected too much. The simulation indicates that these particles are on the left side of the beam before entering the analyzing magnet. By
carefully decreasing the field integral on the inner side of the magnet we might remove this tail without affecting the vertical focusing too much.
We estimate that the effective transport efficiency to the AGOR cyclotron with the modified magnet will increase from $16~\%$ to $45~\%$.

\section{\label{sec:conlook}Conclusions and outlook}
We have performed detailed simulations and measurements of the extraction of a 24~keV He$^+$ beam from an ECR ion source and its subsequent transport through a 110$^{\circ}$ analyzing magnet. We find that such slow ion beams are very sensitive to space-charge forces. Comparing simulations and measurements shows that, for the beam currents investigated (0.5~mA), space-charge compensation is complete and the Coulomb forces between ions can be neglected. The beam emittance behind the extraction system is determined by the strong fringe field of the extraction solenoid of the ECR ion source and amounts to 65~$\pi$~mm~mrad.

The simulations predict significant beam losses on the entrance face of the analyzing magnet and a five-fold increase of the beam emittance during its transport through the magnet. The calculated beam profile and emittance distributions behind the analyzing magnet agree well with measured ones. The simulations clearly show that the emittance blowup is not caused by the initial phase-space distribution of the beam extracted from the ECR ion source, but by the non-paraxiality of the beam leading to kinematic aberrations and by large second-order aberrations of the analyzing magnet. The beam losses can be prevented by increasing the pole gap of the magnet from 67~mm to 110~mm. A better understanding of the kinematic and geometric aberrations of the analyzing magnet has been obtained by expanding its transfer map up to second-order. Using this transfer map together with
detailed 3D magnetic field and ion trajectory calculations we have worked out a proposal to compensate the second-order magnet aberrations by suitably modifying its pole faces. This will result in a significant reduction of the beam losses and emittance growth caused by the
analyzing magnet and will increase the overall transport efficiency to the AGOR cyclotron from 16 to $45\%$.

Current state-of-the-art ECR ion sources are based on superconducting technology to generate both the solenoidal and hexapolar confining fields, and operate with RF-frequencies between 24 and 28~GHz~\cite{lyneis}. The magnetic fields in these sources are thus typically a factor two higher than in the 14~GHz KVI AECR ion source for which the transport properties of the analyzing magnet have been studied in this article. Consequently the emittance of the beams delivered by these sources is also typically a factor two larger than the beam emittance of the KVI-AECR ion source. Compensation of higher-order aberrations in the beam transport caused by both the non-paraxiality of the beams and the higher-order terms in the fields of the optical elements in order to minimize emittance growth and beam losses thus become even more important than it already is in our case. For an optimal design of beam guiding systems for low-energy highly-charged ions a detailed analysis of the higher order aberrations based on realistic 3D fields and of the kinematic perturbations is therefore essential. In addition, since the extracted beam currents are much higher than in our case the problem of space-charge compensation should be given adequate attention in order to fully exploit the potential of these ion sources.

\begin{acknowledgments}
This work is part of the research program of the "Stichting voor Fundamenteel Onderzoek der Materie" (FOM) with financial support from the "Nederlandse organisatie voor Wetenschappelijk Onderzoek" (NWO). It is supported by the University of Groningen and by the "Gesellschaft f\"ur Schwerionenforschung GmbH" (GSI), Darmstadt, Germany.
\end{acknowledgments}

\end{document}